\documentclass[12pt]{iopart}
\usepackage{geometry}                
\usepackage[parfill]{parskip}    
\usepackage{graphicx}
\usepackage{amssymb}
\begin{document}

\title[Fluctuation  mediated interactions due to rigidity mismatch in membranes]{Fluctuation mediated interactions due to rigidity mismatch and their effect on miscibility of lipid mixtures in multicomponent membranes}
\author{David S. Dean}
\address{Universit\'e de  Bordeaux and CNRS, Laboratoire Ondes et Mati\`ere d'Aquitaine (LOMA), UMR 5798, F-33400 Talence, France}
\author{V. Adrian Parsegian}
\address{Department of Physics, University of Massachusetts, Amherst, MA, USA 01003}
\author{Rudolf Podgornik}
\address{Department of Theoretical Physics, J. Stefan Institute, SI-1000 Ljubljana, Slovenia}
\address{Department of Physics, Faculty of Mathematics and Physics, University of Ljubljana, SI-1000 Ljubljana, Slovenia}
\address{Department of Physics, University of Massachusetts, Amherst, MA 01003, USA}

\begin{abstract}
We consider how membrane fluctuations can modify the miscibility of lipid mixtures, that is to say how the phase diagram of a boundary-constrained membrane is modified when the membrane is allowed to fluctuate freely in the case of zero surface tension. In order for fluctuations to have an effect, the different lipid types must have differing Gaussian rigidities. We show, somewhat paradoxically, that fluctuation-induced interactions can be treated approximately in a mean-field type theory. Our calculations predict that,  depending on the difference in bending and Gaussian rigidity of the lipids, membrane fluctuations can either favor or disfavor mixing.   
\end{abstract}
\maketitle
\section{Introduction}
Most biological membranes are composed of a multitude of lipid species. Phase ordering of these lipid species is vital for the biological functions of the membrane \cite{lipids}. Separation of lipids may well be important in biological systems. Indeed, lipids extracted from the erythrocyte membranes form an immiscible two-dimensional liquid, that is very close to the miscibility critical point  \cite{lipmix}. 

Lipid membranes are ideal systems for exploring fluctuation-mediated interactions. Their height fluctuations can be coupled in various manners to their local composition. It has been known for a long time that objects in the membrane, which modify its local mechanical properties, can experience fluctuation-induced forces between them \cite{gou93,stiff,lin11}. In these thermal systems the interactions are energetically of the order of the temperature  $k_BT$. However, in certain circumstances, in particular for tensionless membranes, the interactions can be long-ranged and thus can potentially have an important effect on the organization of the membrane. The Hamiltonian for a tensionless membrane was introduced by Helfrich \cite{helfrich}. In the Monge gauge, appropriate for small fluctuations, the Hamiltonian of the membrane expressed in terms of the membrane height $h({\bf x})$ is given by 
\begin{equation}
H_h = \int d{\bf x} \left({1\over 2} \kappa({\bf x}) \left[\nabla^2 h({\bf x})\right]^2 + \overline{\kappa}({\bf x})\left[{\partial^2h({\bf x})\over \partial x^2} {\partial^2 h({\bf x})\over \partial y^2}-\left({\partial^2h({\bf x})\over \partial x\partial y}\right)^2\right]\right).
\end{equation}
The term $\kappa({\bf x})$ is the local bending rigidity of the membrane and  $\overline{\kappa}({\bf x})$ is the local Gaussian rigidity. The coordinates ${\bf x}=(x,y)$ are the coordinates of the plane representing the projected area of the membrane. In single component membranes, where $\kappa$ and $\overline{\kappa}$ are constant,  the term in the Hamiltonian proportional to $\overline{\kappa}$ is zero when the membrane is  a free-floating sheet. In the Fourier space representation the height-height correlation function in a sheet-like membrane is given by 
\begin{equation}
\langle \tilde h({\bf k}) \tilde h({\bf k}') \rangle = {(2\pi)^2k_BT\delta({\bf k}+{\bf k}') \over \kappa k^4}.
\end{equation}
This means that the bending rigidity can be determined from fluctuation-mode analysis of  tensionless membranes \cite{nig95,strey95} (or in practice membranes with very low tension). The bending rigidity can also be estimated via mechanical measurements, such as by pulling cylindrical tethers from spherical vesicles \cite{bo89}.  Experimental  results show that membrane bending rigidities  typically lie between $3-40k_BT$ \cite{nagle14}. Experimentally estimating the Gaussian rigidity is not  easy  (see \cite{hu12} and the references therein). However, the pure bending energy of a perfectly spherical vesicle, where no fluctuations in the radius occur, can be shown to be given by \cite{boal}
\begin{equation}
H_S  = 4\pi({\overline\kappa}  + 2\kappa).
\end{equation}
The thermodynamic stability of flat membranes obviously requires that $\kappa >0$, to prevent the onset of large unstable fluctuations in one direction while the other remains flat. At the same time stability against formation of vesicles from a flat membrane and against the growth of saddles of mean zero curvature means that 
\begin{equation}
0>{\overline\kappa}  > - 2\kappa.\label{bounds}.
\end{equation}
In reference \cite{hu12} both numerical simulation based and experimental measurements 
for $\overline\kappa$ are reviewed for both bilayer and monolayer systems. The results given for bilayer systems give ${\overline \kappa\over \kappa}\in [-0.9,-0.5]$. For monolayers the results are predominantly such that ${\overline \kappa\over \kappa}\in [-0.9,-0.5]$. It seems reasonable that as the bending rigidities $\kappa$ and $\overline\kappa$ have their origins in the same physical properties of the lipids that they should be of the same order of magnitude.

The first study of fluctuation-induced interactions due to the spatial modulation of bending
rigidities was by Goulian, Bruinsma and Pincus \cite{gou93} who considered the interaction between inclusions, such as proteins, which modify the local rigidities (but do not favor a local mean rigidity). In the case where the rigidity differences due to inclusions are small with respect to the background or mean rigidities (${ \kappa}, {\overline\kappa}_0$), {\sl i.e.}, if we write 
\begin{equation}
\kappa({\bf x}) = \kappa_0 + \Delta\kappa({\bf x})\,;\,  \overline\kappa({\bf x}) = {\overline\kappa}_0 + \Delta{\overline\kappa}({\bf x}),\label{kfluc}
\end{equation}
and assume that $\Delta\kappa$ and $\Delta\overline\kappa$ are small,  one can apply a pairwise approximation, which is exact to second-order in the deviation from the background rigidities, based on the cumulant expansion of the partition function. In \cite{gou93} it was found that the effective two-body interaction between regions deviating from the mean or background rigidities is given by
\begin{equation}
H_2 = {T \over 4\pi^2 \kappa_0^2}\int d{\bf x} d{\bf x'}{ \Delta\overline\kappa({\bf x})\Delta\kappa({\bf x}')
\over |{\bf x}-{\bf x}'|^4}.\label{H2}
\end{equation}
For regions (say discs) whose centers of area are separated by a distance $R$ and are of area  $S_1$ and $S_2$ respectively, we see that when $R$ is much larger than the size of the regions, the first-order term in a multipole expansion of the energy between the two 
regions is given by
\begin{equation}
 H_2 = {T S_1S_2(\Delta\kappa_1\Delta\overline\kappa_2 + \Delta\kappa_2\Delta\overline\kappa_1 )\over 4\pi^2 \kappa_0^2 R^4}.
\end{equation}
We first notice that this is a  long-range interaction, and it is clearly a fluctuation-induced interaction which can be inferred  from its proportionality to $T$. Secondly, we see that at the pairwise-order, we see that one must have both variations in $\kappa({\bf x})$ and $\overline\kappa({\bf x})$ in order to have an interaction.
In the case where both inclusions are of the same type,  so that $\Delta \kappa_1 =\Delta\kappa_2=\Delta\kappa$ and   $\Delta \overline\kappa_1 =\Delta\overline \kappa_2=\Delta\overline\kappa$, we see that if 
$\Delta\kappa$ and $\Delta \overline \kappa$ have the same sign, then the interaction is repulsive, where as if they have opposite signs the interaction is attractive. This is an intriguing result. To to date, other than via direct calculation, no one has proposed a physical explanation for the sign of the interaction.

The same problem can also be analyzed for stiff inclusions and in general in the limit where the variations in the rigidity are not small \cite{stiff}. Stiff inclusions can be modeled by imposing the condition that membrane  be locally flat for a variety of objects, such as discs and rods. Inclusions which modify the rigidity in a point-wise manner, {\sl via} the delta-function-like changes to the rigidity, can be analyzed exactly and in principle for any number of objects. All of these studies consistently confirm the long-range interaction predicted in \cite{gou93}. More recently, methods developed for the study of the quantum electromagnetic Casimir effect based on a scattering matrix approach have been employed to examine the interaction between two discs in a membrane \cite{lin11}. Within this formalism, all n-body effects between the two discs can be taken into account; in the tensionless limit, it is found that the interaction between two discs of radius $a$ behaves, for large separations $R$ between the disc centers,  as
\begin{equation}
H_2= -T A {a^4\over R^4}
\end{equation}
where the coefficient $A$ is given by
\begin{equation}
A = 4{\overline \kappa_0 -\overline{\kappa}\over 4\kappa_0 +\overline\kappa_0-
\overline\kappa}\left({\overline \kappa_0 -\overline{\kappa}\over 4\kappa_0 +\overline\kappa_0-
\overline\kappa} +{\kappa -\kappa_0+{1\over 2}(\overline\kappa-\overline\kappa_0)\over 2\kappa +\overline\kappa-\overline\kappa_0}\right),\label{2bexact}
\end{equation}
where $\kappa$ and $\overline{\kappa}$ are the rigidities of the disc and $\kappa_0$ and $\overline{\kappa}_0$ are the background rigidities of the surrounding membrane. We see again that, in order to have an interaction, a variation in the Gaussian curvature is necessary. However, at this higher order the interaction persists even if $\kappa$ is constant.

The computation of the fluctuation-mediated interaction in this simple two-body system is on the face of it rather complicated. Practically, apart from two-body interactions, one would prefer to know how to describe, {\sl e.g.}, phase diagrams for many-particle systems, specified by macroscopic quantities, such as their average density, interacting via fluctuation-induced interactions. One example would be to examine a membrane containing a finite density of  proteins that locally modify the bending rigidities. Another would be a model where the membrane is composed of two lipid species that have different rigidities, which is the case we will study below. 

Few studies exist on the thermodynamics of systems with many inclusions \cite{netz95,weik01,dea06}, though it has been shown in both Monte Carlo simulations and using various approximation schemes, such as mean-field theory coupled with Monte Carlo simulations and cumulant expansions, that fluctuation-induced interactions can have a significant influence on the organization of inclusions in lipid membranes.  The studies described in  \cite{netz95,weik01,dea06} are, however, quite different from that expounded here. First, we consider a free-floating tensionless membranes that are not subject to external potentials, such as a harmonic confining potential \cite{netz95}, or to an imposed external pressure \cite{weik01}. In \cite{dea06} the effect of variations in bending rigidity for membranes under tension was considered, but no variation in Gaussian rigidity was included. Secondly, the underlying formal analysis is also quite different from these previous studies. As mentioned above, the simple example we will consider here is that of lipid mixtures where the different lipid components have different rigidities. In principle this problem is complicated by the fact that each leaflet of the membrane can be composed of different lipid types. In this study we will assume that the lipids on both leaflets are the same. However, the method of analysis we propose here could be readily adapted to a model for a genuine two-leaflet system, notably because of its simplicity in application. 

The most easily applicable theory to analyze phase transitions is the mean-field theory. While it has several quantitative failings in critical systems, it is a useful tool to determine phase diagrams and is the first choice of  analysis  in most  problems. It, however, appears futile to apply mean-field theory to fluctuation-induced interactions as in such systems there is no mean field; for instance variations of  the rigidities in the Helfrich Hamiltonian clearly do not break the up/down symmetry of the membrane. In what follows we will demonstrate that the Helfrich theory can in fact be written in such a way as to allow the formulation of a mean-field theory that does capture fluctuation-induced interactions and, moreover, in such a way that the pairwise result of \cite{gou93} is perfectly taken into account. We will show how this mean-field theory modifies the standard regular solution mean-field theory \`a la Flory describing demixing transitions. 

In addition, it is well known that Casimir-like fluctuation-induced interactions  often lead to divergent free energies that need to be regularized by introducing an ultra-violet or short distance cut-off. For example, in the electromagnetic Casimir effect there are surface and bulk divergent terms in the energy, that nevertheless do not contribute to the Casimir force if the bodies keep the same form and composition. In the theory we develop, we need to define a membrane patch size that is of the order of a lipid size and to specify the composition of the patch in terms of the lipid type occupying the patch. At the same time, this patch size corresponds to a lattice spacing for the Helfrich elastic Hamiltonian and thus plays the role of the natural cut-off for the membrane fluctuations. Therefore, both the underlying lattice model for the lipid composition of the system and the lattice on which the membrane fluctuations take place are the same. This means that the cut-off will only set an overall energy scale and the phase diagram will thus be cut-off independent.

The paper is organized as follows. In section (\ref{pw}) we discuss fluctuation-induced interactions in the pairwise approximation and also show that the problem of tensionless membranes with constant Gaussian curvature can be solved exactly. Then in section (\ref{mf})  we discuss, not only how a mean-field theory for fluctuation-induced interactions can be formulated but also demonstrate the pitfalls associated with the most naive mean-field theory. We show how the correct pairwise interaction physics can be implemented at a mean-field level by changing the variables of the field theory and making it effectively non-local. This reformulation of the theory has two main advantages: it not only means that the  mean-field theory captures the two body interactions correctly, but it also explains why differences in the Gaussian rigidity are necessary to generate interactions. This latter point has been understood in slightly different contexts by a number of authors \cite{netz95,weik01,kac13}, but we revisit it here as it is of vital importance in constructing the theory and may also have experimental consequences.  In section (\ref{mfl}) we show how the mean-field analysis of the membrane fluctuations can be coupled with the standard mean-field theory known as {\em regular solution theory}, in a simple lattice based model, to see how the mean-field phase diagram is modified by membrane fluctuations. In other words, we see how the phase diagram of a perfectly flat membrane, adhered to a flat surface or held under tension in a frame, is modified if it is allowed to fluctuate under zero tension. We discuss the predicted modification of the phase diagram and notably the effect of height fluctuations on mixing-demixing temperatures. The underlying mean-field theory is then resummed by formulating it variationally in section (\ref{varimpro}). Though the basic results are unchanged at the two-body level, this resummation predicts subtle higher order differences from the basic mean-field theory, notably the presence of interactions in the case where the bending rigidity is constant. Finally, we discuss possible experimental verification of our predictions and directions for further study.
  
\section{Pairwise approximation and exact results for fluctuation-induced interactions}\label{pw}

There are many examples of fluctuation-induced interactions. Such interactions are generated between objects that interact with or modify the fluctuations of a quantum or thermal field \cite{gol99}. The most important and best known of these interactions are  van der Waals  forces, which in the appropriate limit yield the celebrated Casimir force \cite{par06}. Both dielectrics and conductors are coupled to the electromagnetic field; however, they do not break the symmetry of the field as charges would. Mathematically, and indeed physically, their effect can be taken into account via a quadratic coupling to the electromagnetic field. However, this quadratic coupling means that the naive mean-field of the theory, obtained from taking the saddle point of the Hamiltonian, is zero.

We begin by considering the case of a tensionless membrane but where the Gaussian bending rigidity is constant and thus does not contribute to the elastic energy. We thus have
\begin{equation} 
H_{Hel}(\Delta\overline\kappa=0) = {1\over 2}\int d{\bf x} \kappa({\bf x})\left[\nabla^2 h({ \bf x})\right]^2. 
\end{equation}
If we consider the case where the variations of $\kappa$ with respect to the background value $\kappa_0$ are small, we can compute the partition function due to elastic fluctuations using the cumulant expansion as in \cite{gou93}. The one-body terms are  independent of the arrangement of the particles responsible for the variation of $\kappa$, the effective two body interaction is, however, given by
\begin{equation}
H_2(\Delta\overline\kappa=0) = -{T\over 4}\int d{\bf x} d{\bf x}'
\Delta\kappa({ \bf x})\Delta\kappa({\bf x}')\left[\nabla^4 G_{Hel}({\bf x}-{\bf x'})\right]^2,
\end{equation}
where $G_{Hel}$ is the Green's function obeying
\begin{equation}
-\kappa_0 \nabla^4 G_{Hel}({\bf x}-{\bf x}') = -\delta({\bf x}-{\bf x}').
\end{equation}
This gives
\begin{equation}
\fl H_2(\Delta\overline\kappa=0) = -{T\over 4\kappa_0^2}\delta({\bf 0})\int d{\bf x} d{\bf x'}
\Delta\kappa({\bf x})\Delta\kappa({\bf x}')\delta({\bf x}-{\bf x}')= -{T\over 4\kappa_0^2}\delta({\bf 0})\int d{\bf x} 
\Delta\kappa({\bf x})^2,
\end{equation}
which is  a zero-range interaction and does not change the equilibrium configurations of the particles if they are not permitted to overlap.  

We can demonstrate this lack of interaction at all orders by changing variables in the partition function. If we define a new variables $u({ \bf x}) =- \nabla^2 h({ \bf x})$ ( minus the local average curvature) the resulting Helfrich partition function, up to a Jacobian factor which is independent of the rigidity, becomes 
\begin{equation}
Z_{Hel}(\Delta\overline\kappa=0)=\int d[u]\exp\left(-{\beta\over 2}\int \kappa({\bf x})u^2({\bf x}) d{\bf x}\right),
\end{equation}
Now making another change of variable $u({\bf x}) = w({\bf x})/\sqrt{\beta\kappa({\bf x})}$, we find that the contribution to the Helfrich free energy from particle configurations is given by
\begin{equation}
F_{Hel}(\Delta\overline\kappa=0) = {T\over 2}\sum_{\bf x} \ln\left(\beta\kappa({\bf x})\right) \to {T\over 2a^2}\int \ln\left(\beta\kappa({\bf x})\right)d{\bf x},
\end{equation}
where we have evaluated the functional integral on a lattice of spacing $a$ and then taken the continuum limit. This result tallies with the contribution of electromagnetic field fluctuations to all orders in n-body interactions within a mean-field
theory \cite{dea12}.

For a mixture of lipid types 1 and 2 with volume fractions $\phi$ and $1-\phi$ and bending rigidities $\kappa_1$ and $\kappa_2$ respectively, we then find
\begin{equation}
F_{Hel}(\Delta\overline\kappa=0) = {NT\over 2}\left[\phi\ln(\beta\kappa_1) + (1-\phi)\ln(\beta\kappa_2)\right]\label{kb0}
\end{equation}
where $N = A/a^2$ is the number of independent membrane patches for a two-dimensional  membrane of projected area $A$. Note that from this free energy we find that the internal energy of the system is given by
\begin{equation}
U = {N T\over 2},
\end{equation}
that is to say the energy of $N$ membrane patches with an underlying quadratic Hamiltonian as expected from the equipartition of energy. Thus again we see that there is no interaction. This has been pointed in \cite{netz95,weik01} for the case  of membranes and in \cite{kac13} for the case of semi-flexible polymers. The point here is that the physically relevant variable is the mean local curvature and that it is statistically independent point by point. There is, however, an additional subtle point, if the  field $h$ has boundary conditions, then the change of variables made is not strictly valid. If we have free boundaries but with a line tension, the total length of the membranes perimeter will depend on the position of the particles, and thus the interaction induced by  rigidity variations in the absence of surface tension will give a sub-extensive change in the free energy proportional to the perimeter of the membrane. We thus see that, in the absence of surface tension and variations in the Gaussian rigidity, the membrane fluctuations do not induce interactions between regions of different bending rigidity. 

It is also interesting to note that one can compute the height-height correlation function for
the membrane in this tensionless constant Gaussian rigidity case. In terms of the variable 
$w$ the height is given by
\begin{equation}
h({\bf x}) = \int G({\bf x}-{\bf x}''){w({\bf x}'')\over \sqrt{\beta \kappa({\bf x}'')}} d{\bf x}''
\end{equation}
where 
\begin{equation}
\nabla^2 G({\bf x}-{\bf x}") = -\delta({\bf x}-{\bf x}').\label{g1}
\end{equation}
The measure on the field $w$ is then simply given by
\begin{equation}
P[w] = {\exp\left(-{1\over 2}\int w^2({\bf x}) d{\bf x}\right)\over \int d[w]\exp\left(-{1\over 2}\int w^2({\bf x}) d{\bf x}\right)}.
\end{equation}
From this it is easy to see that
\begin{equation}
\langle h({\bf x})h({\bf x}')\rangle =T\int d{\bf x}'' {G({\bf x}-{\bf x}'') G({\bf x}''-{\bf x}') \over \kappa({\bf x}'')}.
\end{equation}
Consequently in a statistically translationally invariant system the spatially averaged correlation function is given by
\begin{equation}
\overline{\langle h({\bf x})h({\bf x}')\rangle} =T\int d{\bf x}'' {G({\bf x}-{\bf x}'') G({\bf x}''-{\bf x}') \over \kappa_e}.
\end{equation}
where 
\begin{equation}
\kappa_e = \langle {1\over \kappa}\rangle^{-1} \label{kehm};
\end{equation}
the effective bending rigidity is then given by the harmonic mean. We note that Jensen's inequality 
implies that $\kappa_e \le \langle \kappa\rangle $, thus the membrane is softened with respect to a pure one with $\kappa=\langle \kappa\rangle$, the arithmetic mean of the rigidities. Equivalently for small $k$ in Fourier space we rewrite the height correlator as
\begin{equation}
\overline{\langle \tilde h({\bf k}) \tilde h({\bf k}') \rangle} = {(2\pi)^2T\delta({\bf k}+{\bf k}') \over \kappa_e k^4}.
\end{equation}

The question of the effective bending rigidity for a model with constant $\overline\kappa$ at zero tension but in a quadratic confining potential was addressed by Netz and Pincus \cite{netz95}. In their cumulant expansion they perturbatively computed the effective bending rigidity for a quenched distribution of fluctuations of $\kappa({\bf x})$;  their perturbative result can be rewritten as the harmonic mean. The formula Eq. (\ref{kehm}) has also been proposed in \cite{keos} by invoking more phenomenological arguments. The agreement here is logical as the distribution of $\kappa$ is effectively decoupled from the membrane fluctuations. The result that $\kappa_e$ is the harmonic mean is a zero order result. When fluctuations in $\overline\kappa$, a non-zero surface tension or quadratic coupling are included the value of $\kappa_e$ will be suitably renormalized.  

\section{mean-field theory for fluctuation-induced interactions}\label{mf}

We have seen that the Helfrich Hamiltonian $H_{Hel}$ depends on the local configuration of the membrane components via the functions $\kappa({\bf x})$ and  $\overline\kappa({\bf x})$. The total Hamiltonian on a lattice will have two components, a direct interaction between the membrane components $H_D$ plus the membrane elasticity term.  We thus  write the total Hamiltonian as
\begin{equation}
H_T = H_D + H_{Hel}.
\end{equation}
The Hamiltonian for $H_D$ is that of a lattice gas and depends on the occupation number of, say, membrane or lipid type $k$ at the site $i$. We can, for instance, define the variable $n_i$ which is equal to $1$ at site $i$ if the lipid is of type 1 and $0$ if it is of type 2. The lipids of different types will have rigidities denoted by $\kappa_i$ and $\overline\kappa_i$.  This model can, of course, be generalized to any number of lipid types. The simplest mean-field approximation one can make is to use a non-interacting lattice gas as the trial Hamiltonian for the particles on the lattice. We have then the total partition function  given by
\begin{equation}
Z_T = {\rm Tr}\int d[h] \exp(-\beta H_D -\beta H_{Hel}),
\end{equation}
where ${\rm Tr}$ denotes the sum over the particle configurations on the lattice.

The mean-field approximation to this partition function is given by
\begin{equation}
Z_{MF} =\int d[h] Z_0 \exp\left(-\beta \langle H_D-H_0\rangle_0 -\beta \langle H_{Hel}\rangle_0 \right),
\end{equation}
where $\langle \cdot\rangle_0$ indicates the average with respect to the non-interacting lattice gas and $Z_0$ is the partition function for the lattice gas. The mean-field partition function also bounds the exact partition function from below and thus the mean-field free energy gives an upper bound for the free energy. We then remain with
\begin{equation}
Z_{MF} = \exp(-\beta F_{MFD})\int d[h]\exp\left(-\beta \langle H_{Hel}\rangle_0 \right),
\end{equation}
where $F_{MFD}$ is the mean-field free energy for the system without height fluctuations (MFD signifying mean-field-direct for the direct interactions in the lattice model). It will have the form
\begin{equation}
F_{MFD}= N f_{mfd}(\phi)
\end{equation}
where $N$ is the number of lattice sites. For example, for a symmetric binary mixture undergoing a continuous demixing transition, regular solution theory has a free energy per lattice site given by \cite{gas08}
\begin{equation}
\beta f_{mfd(\phi)} = \chi_d \phi(1-\phi)+   \phi\ln(\phi) + (1-\phi)\ln(1-\phi)
\end{equation}
where $\phi$ and $1-\phi$ are respectively the factions of lipids of type 1 and 2, and $\chi_d$ is the Flory parameter measured in units of $T$. When $\chi_d > 0$ the interaction favors demixing of the system. The free energy $F_{MFD}$ thus describes the mean-field free energy of a confined system which is not allowed to fluctuate. This could be achieved for instance by applying a large lateral tension that generates an effective surface tension which suppresses all fluctuations.  

It now remains to compute the term $ \langle H_{Hel}\rangle_0$. However, this mean-field approximation is only accurate to first order in the cumulant expansion, and we know that fluctuation-induced interactions only appear at second order. A  mean-field approximation thus appears to be rather hopeless. Furthermore, if we use this naive mean-field approximation, we find that the membrane contribution to the total free energy (the membrane mean-field - MMF-
free energy) is, up to a constant independent of its composition, given by
\begin{equation}
\fl F_{MMF}=-T\ln\left[\int d[h]\exp\left(-\beta \langle H_{Hel}\rangle_0 \right) \right]= {TN\over 2}\ln(\langle\kappa\rangle_0)={TN\over 2}\ln\left(\phi\kappa_1 +(1-\phi)\kappa_2\right),
\end{equation}
where we have carried out the functional integral in Fourier space with the lattice cut-off $-{\pi\over a}<k_x,\ k_y<{\pi\over a}$, and note that the projected area $A$ and $N$ are related by $N= A/a^2$. The term proportional to $\overline\kappa$ gives zero upon averaging and thus the result is independent of $\overline\kappa$. This is clearly an undesirable feature. Our analysis in section (\ref{pw}) shows that variations in $\overline\kappa$ are essential to induce fluctuation interactions.  Furthermore, it is straightforward to see that when $\overline\kappa$ is constant, the mean-field approximation predicts a spurious tendency of membrane fluctuations to favor demixing, while our exact result shows that, in this case, membrane fluctuations play no role in how the membrane is organized. 

Naive mean-field theory applied to this problem is thus incapable of capturing fluctuation-induced interactions even at the pairwise level and in addition introduces an artifactual tendency toward demixing that we know is not present in the case where $\overline\kappa$ is constant. The solution to this problem is, as in section (\ref{pw}), to express the membrane partition function in terms of the variable $w({\bf x}) = -\sqrt{\beta\kappa({\bf x})}\nabla^2h$. As we have seen already, this change of variables does not change the partition function as a function of its composition as long as the overall composition is fixed. With this change of variables the membrane partition function is given by
\begin{eqnarray}
\fl Z_M=Z_{Hel}(\Delta\overline\kappa=0) \int d[w]\exp\left(-{1 \over 2}\int d{\bf x} w^2({\bf x}) \right.\nonumber\\
\fl \left.+ \int d{\bf x} d{\bf x}' d{\bf x}''\overline{\kappa}({\bf x})
\left[{\partial^2G({\bf x}-{\bf x}')\over \partial x^2} {\partial^2 G({\bf x}-{\bf x}'')\over \partial y^2}-{\partial^2G({\bf x}-{\bf x}')\over \partial x\partial y}{\partial^2G({\bf x}-{\bf x}'')\over \partial x\partial y}\right]{w({\bf x}')w({\bf x}'')\over \sqrt{\kappa({\bf x})}\sqrt{\kappa({\bf x}')}}\right),\nonumber \ \label{zm2}
\end{eqnarray}
where $G$ is the Green's function defined in Eq. (\ref{g1}) and the first term is that for a membrane with no differences in Gaussian rigidity coming from the change of variables. Now if we write the rigidities as small fluctuations about a background field as in Eq, (\ref{kfluc}) and take just the first term in the cumulant expansion of $Z_M$ written in the form  of Eq. (\ref{zm2}), we find that 
\begin{equation}
Z_M = \exp(-\beta H_2),
\end{equation}
where the Hamiltonian $H_2$ depends on the rigidity $\kappa$ at two points, given exactly by Eq. (\ref{H2}). Therefore by reformulating the problem we have found a representation for the membrane partition function that contains the second-order cumulant expansion of the original representation in terms of the height variable $h$ in its first-order cumulant expansion. The use of mean-field theory in this representation is thus clearly superior. In addition, the mean-field result when $\overline \kappa$ is constant also agrees with the corresponding exact result available for this case. Note that in principle the second order term in the cumulant expansion in this representation could contain a pairwise interaction term of order $\Delta\overline \kappa^2$. However one can directly check that this term is zero, as should  be the case.

\section{Mean-field theory on a lattice}\label{mfl}

Here we consider the mean-field theory for tensionless membranes regularized on a lattice. First we consider the basic formulation of mean-field theory using the representation that accounts for two body fluctuation-induced interactions. Secondly we resum this basic result using a variational reformulation of the problem, exploiting the solvability of the model for any rigidity field $\kappa({\bf x})$ in the absence of spatial variations of $\overline\kappa({\bf x})$. 

\subsection{Basic mean-field theory}
In this section we develop the mean-field theory suggested in section (\ref{mf}) for a system  which is regularized by placing it on a lattice. We will take a square lattice with lattice spacing $a$. The lattice membrane Hamiltonian can be expressed in terms of the discrete operators $D_x$ and $D_y$ defined by
\begin{equation}
\fl D_x f({\bf x}) ={1 \over 2 a}[f(x+a,y)-f(x-a,y)]\ ; \  D_y f({\bf x})={1 \over 2 a}[f(x,y+a)-f(x,y-a)],
\end{equation}
and the lattice Laplacian
\begin{equation}
\Delta = D_x^2 + D_y^2.
\end{equation}
The membrane Hamiltonian is then given by
\begin{equation}
H_{Hel} ={a^2\over 2} \sum_{\bf x} \kappa({\bf x}) [\Delta h]^2 + {a^2} \sum_{\bf x}
\overline{\kappa}({\bf x})(D^2_x h D^2_y h - [D_xD_y h]^2).
\end{equation}
 As in the continuous case, we can rigorously perform the change of variables 
 $u({\bf x}) = -\Delta h$ then $w({\bf x}) = \sqrt{\beta\kappa({\bf x})}u({\bf x})$.
 to obtain a Hamiltonian in terms of the variable $w$ that is given by
 \begin{eqnarray}
\fl \beta H'_{Hel} = {a^2\over 2}\sum_{\bf x} w^2({\bf x}) + \nonumber \\ \fl a^2 \sum_{{\bf x},{\bf x}',{\bf x}''}\overline\kappa({\bf x})\left[D_x^2 G_L({\bf x},{\bf x}') D_y^2 G_L({\bf x},{\bf x}'')  -
D_xD_y G_L({\bf x},{\bf x}') D_xD_y G_L({\bf x},{\bf x}'')\right]{w({\bf x}')w({\bf x}'')\over \sqrt{\kappa({\bf x}')}\sqrt{\kappa({\bf x}'')}},\nonumber \\
\end{eqnarray}
where $G_L$ is the lattice Green's function obeying 
\begin{equation}
\Delta G({\bf x}, {\bf x}')=-\delta_{{\bf x},{\bf x}'}.
\end{equation}
From the Fourier representation of the lattice Green's function \cite{itdr} it is easy to show
that at coinciding points
\begin{equation}
\fl D^2_x G({\bf x},{\bf x})  =-{1\over 2}\ ; \ D^2_y G({\bf x},{\bf x})=-{1\over 2}\ ;\ D_xD_yG({\bf x},{\bf x})=D_yD_xG({\bf x},{\bf x})=0.\label{god}
\end{equation}
Now the mean-field approximation requires the computation of $\langle \beta H'_{Hel}\rangle_0$, the average with respect to the non-interacting lattice gas Hamiltonian. To do this we  note that three-point  correlation functions for the free lattice gas depend
on whether or not the spatial points in the average coincide or not; thus we have the 
general expression
\begin{equation}
\langle \overline\kappa({\bf x})\kappa({\bf x}')\kappa({\bf x}'')\rangle_0
=\delta_{{\bf x}{\bf x}'}\delta_{{\bf x},{\bf x}''}\alpha + (\delta_{{\bf x}{\bf x}'}+\delta_{{\bf x}{\bf x}''})\beta + \delta_{{\bf x}'{\bf x}''}\gamma + \delta,
\end{equation}
where 
\begin{equation}
\alpha = A-2B -C  + 2D;\  \beta = B-D;\  \gamma = C-D; \ \delta = D
\end{equation}
with 
\begin{equation}
A = \langle {\overline\kappa\over \kappa}\rangle_0;\ B = \langle {\overline\kappa\over \sqrt{\kappa}}\rangle_0\langle {1\over \sqrt{\kappa}}\rangle_0;\ C = \langle \overline\kappa\rangle_0\langle {1\over {\kappa}}\rangle_0; \ D = \langle \overline\kappa\rangle_0\langle {1\over \sqrt{\kappa}}\rangle^2_0.
\end{equation}
Terms which contract the coordinate ${\bf x}$ with one of the others are zero (such terms are present for $\overline\kappa$ constant and so are zero). We now use the  formulas in Eq. (\ref{god}) to simplify the remaining terms to find
\begin{equation}
\langle \beta H'_{Hel}\rangle_0 = {1\over 2} \sum_{{\bf x}} w^2({\bf x}) ( 1+ {\alpha + 2\beta\over 2}).
\end{equation}
The integral over the variables $w({\bf x})$ is then straightforward to compute. We find that the part of the mean-field free energy due to fluctuations and depending on the composition is given by
\begin{equation}
F_{MMF} = {TN\over 2}\left(
\ln\left(1+ {1\over 2} \left[{\langle {\overline\kappa\over \kappa}\rangle}_0-\langle \overline\kappa\rangle_0\langle {1\over {\kappa}}\rangle_0\right] \right) + \langle\ln(\beta \kappa)
\rangle_0\right).\label{mmf1}
\end{equation}
Note that the second term above, stemming from the change of variables, is the free energy for the same system but with equal Gaussian rigidities as given in Eq. (\ref{kb0}) and does not include any interaction between different regions. For the two component system considered here (the result of course can be generalized to any number of components), we find that
\begin{equation}
\langle {\overline\kappa\over \kappa}\rangle_0-\langle \overline\kappa\rangle_0\langle {1\over {\kappa}}\rangle_0 = (\overline\kappa_1 -\overline\kappa_2)({1\over \kappa_1}-{1\over \kappa_2})\phi(1-\phi).
\end{equation}
We thus find a membrane contribution to the free energy per site  given by
\begin{equation}
\beta f_{mmf}(\phi) = {1\over 2}\ln\left(1+ 2\chi_f\phi(1-\phi)\right) + 
\phi\ln(\beta\kappa_1) + (1-\phi)\ln(\beta \kappa_2)\label{mfm},
\end{equation}
where $\chi_f$ is an effective Flory parameter induced by membrane fluctuations given by
\begin{equation}
\chi_f = {1\over 4}(\overline\kappa_1 -\overline\kappa_2)({1\over \kappa_1}-{1\over \kappa_2}).\label{echif}
\end{equation}
When $\chi_f$ is positive the effect of fluctuations is to favor demixing. This consistent with the observation in the pairwise approximation that if $\Delta\overline\kappa$ and $\Delta\kappa$ have the same sign, then the interaction is repulsive and mixing is thus thermodynamically favored by the composition coupling to the height fluctuations. In this case the parameter $\chi_f$ is negative and thus mixing is also favored in the mean-field theory. 

In systems where lipids have large rigidity mismatches,  the bending and Gaussian rigidities for a given lipid type should have the same order of magnitude, we should thus  expect that  $\chi_f >0$ (bearing in mind that the $\overline \kappa$ are negative) and membrane fluctuations should favor demixing, consequently raising the demixing temperature of lipids with very different rigidities. In the case where $\overline\kappa_i=-2\kappa_i$ for $i=1,\ 2$, i.e. we have the maximal value of $|\overline \kappa|$ for  both species of lipid, we find that 
\begin{equation}
\chi_f = {1\over 2\kappa_1\kappa_2}(\kappa_1 -\kappa_2)^2.
\end{equation}
and thus an effective interaction favoring demixing. We note that the mean-field theory predicts that  differences in both $\overline \kappa$ and  $\kappa$ are necessary to have an effective fluctuation-induced interaction, in agreement with the pairwise calculation in Eq. (\ref{H2}).  Furthermore the membrane mean-field free energy is always finite as the bounds in Eq. (\ref{bounds}) ensure the inequality $\chi_f >-1/2$. The full n-body result for two discs of \cite{lin11}, clearly shows that interactions should occur even when $\kappa$ is constant, these are higher body effects which are missed by the first term in the cumulant expansion and  and hence by our basic mean-field calculation.

If one includes the contribution to the mean-field free energy from the underlying lattice model from the mean-field regular solution theory, the total mean-field free energy is obtained as
\begin{equation}
\fl \beta f_t = {1\over 2}\ln\left(1+ 2\chi_f\phi(1-\phi)\right)+\chi_d \phi(1-\phi)+   \phi\ln(\phi) + (1-\phi)\ln(1-\phi) + \phi\ln(\beta\kappa_1) + (1-\phi)\ln(\beta \kappa_2).\label{regsf}
\end{equation}
When the function $f$ becomes concave the mean-field approximation is interpreted as a thermodynamic instability leading to demixing into two phases, the compositions of which  are determined via the tangent construction and the lever rule. Notice that the non-interacting term for $\Delta\overline\kappa=0$ which is written as the last term of Eq. (\ref{regsf}) is linear in $\phi$ and plays no role in demixing. When the direct interaction  $\chi_d$ is taken to zero so that the only interactions present are due to height fluctuations it is straightforward to see that the free energy is always convex and no demixing can occur. Strictly speaking only the first-order term in $\chi_f$ is exact. When this term alone is taken into account a demixing transition is possible when the total or effective Flory parameter $\chi_t = \chi_f+\chi_d > 2$. Even when $\chi_d=0$ this in equality can be achieved. However, our resummed result (leading to the logarithm in the first term of Eq. (\ref{regsf})  suggests that n-body interactions have the effect of frustrating the attraction between similar lipid types and reducing the interaction with respect to that expected from the two body interaction (an effect reminiscent of the saturation of van der Waals forces at high dielectric contrasts \cite{par06}).

We can estimate the importance of fluctuation-induced interactions in this mean-field theory by estimating the shift in the critical temperature $T^{(0)}_c$ when fluctuation-induced interactions are included. Consider the following case, motivated by the data given in \cite{hu12}, where $\overline{\kappa_i} = -\kappa_i$ and where take as an example $\kappa_2=2\kappa_1$; yielding $\chi_f=1/16$. From this we find that a free-floating membrane has a demixing temperature $T_c\approx 1.0312 T^{(0)}_c$. For a transition temperature of $303$K this corresponds to an increase of $9$K in the demixing transition temperature.

For systems where $\chi_d >0$ where phase separation can occur, the contribution of height fluctuations  raises the transition temperature when $\chi_f>0$ and lowers the  transition temperature when $\chi_f<0$. In general, one should expect that $\kappa$ and $\overline\kappa$ for a single lipid species should be of the same order of magnitude since the energy scales of these respective bending energies are determined at a molecular level, indeed it should be noted that in many measurements it is found that $\overline\kappa\sim -\kappa$. This means that for lipids with a large bending rigidity mismatch, say $\kappa_1\gg \kappa_2$, we should expect that $|\overline\kappa_1|\gg|\overline\kappa_2|$ and thus  we should expect that $\chi_f> 0$. That is to say that mismatched lipids should have a tendency to demix due to membrane height fluctuations in tensionless or near tensionless membranes.

Using the results derived above, it is straightforward to compute the first nontrivial correction to the effective rigidity. We  find that
\begin{equation}
\fl \kappa_e = \langle {1\over\kappa}\rangle_0^{-1}\left[1 + {1\over 2} \langle {\overline \kappa \over \kappa}\rangle_0 -{1\over 2} \langle \overline\kappa\rangle_0 \langle {1 \over \kappa}\rangle_0\right]={\kappa_1 \kappa_2\over \phi\kappa_2 + (1-\phi)\kappa_1}\left[1 + 2\phi(1-\phi)\chi_f\right].
\end{equation}
Deviations of $\kappa_e$ from the harmonic mean  bending energy given in Eq. (\ref{kehm}) thus indicate a difference in the Gaussian rigidity of the lipids and that  this difference can thus be estimated via the expression for $\chi_f$ given in Eq. (\ref{echif}). Interestingly if $\chi_f$ is negative, and thus favors mixing, the effective rigidity is reduced. This increased tendency toward mixing induced by height fluctuations feeds back to soften the membrane.

\begin{figure}[t]
   \centering
  \includegraphics[width=14cm]{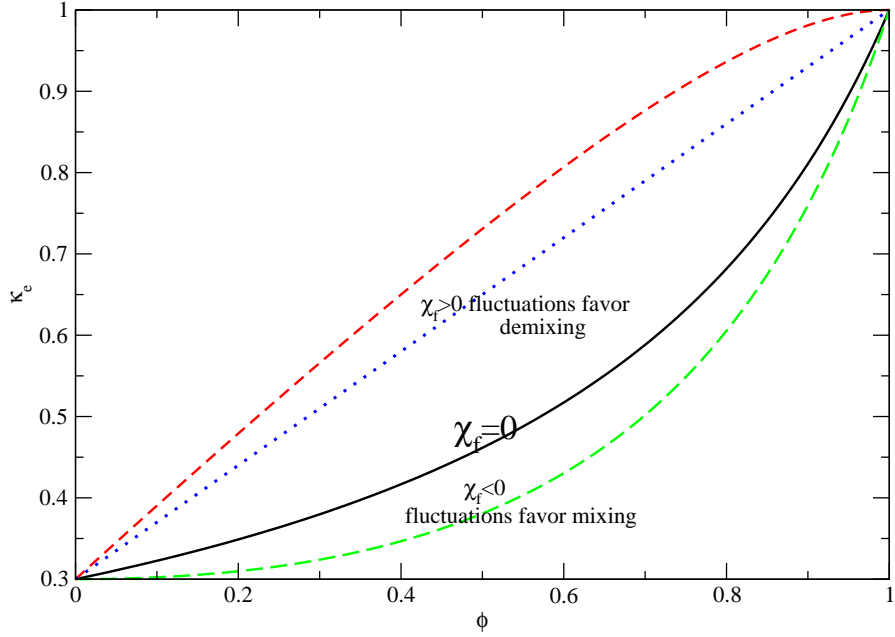}
  \caption{ Basic mean-field theory prediction for the effective bending rigidity of a membrane in terms of its composition $\phi$ with $\kappa_1=1$ and $\kappa_2=0.3$
The solid black curve is the case where there are no interactions corresponding to $\chi_f=0$. The upper dashed curve (red -color on line) corresponds to the limiting case where $\overline\kappa_1=-2\kappa_1=-2$ while $\overline\kappa_2=0$. The upper 
dashed curve (green - color on line) corresponds to the other limiting case where $\overline\kappa_2=-2\kappa_2=-0.6$ while $\overline\kappa_1=0$. The dotted (blue- color online)  line corresponds to the case where $\overline\kappa=-2\kappa$ for both species.}
\label{kappabas}
\end{figure}

In Fig. (\ref{kappabas}) we show the form predicted for the effective bending rigidity $\kappa_e$ as a function of $\phi$. We have chosen the case where in the appropriate units we have $\kappa_1=1$ and $\kappa_2=0.3$.  The solid black line in the middle corresponds to the interactionless case $\chi_f=0$ and thus the absence of fluctuation-induced interactions When $\chi_f<0$ and fluctuations thus favor mixing, the curve lies below that of the interaction less one, indicating softening and when $\chi_f>0$ the opposite occurs. The two dashed curves present the limiting cases $\overline \kappa_{1}=-2\kappa_{1}$ with $\overline \kappa_{2}=0$ (top curve) and $\overline \kappa_{2}=-2\kappa_{2}$ with $\overline \kappa_{1}=0$ (bottom curve), where the bounds of Eq. (\ref{bounds}) are saturated and we have the maximal values for $|\chi_f|$.

\subsection{Variationally improved mean-field theory}\label{varimpro}

The calculation carried out above can be modified by carrying out a resummation of the basic theory. Firstly we trivially rewrite the Helfrich Hamiltonian as
\begin{equation}
\fl H_{Hel} ={a^2\over 2} \sum_{\bf x} (\kappa({\bf x})+\lambda({\bf x})) [\Delta h]^2 + {a^2} \sum_{\bf x} \overline{\kappa}({\bf x})(D^2_x h D^2_y h - [D_xD_y h]^2)-{\lambda({\bf x})\over 2}[\Delta h]^2,
\end{equation}
where $\lambda({\bf x}) $ is a function that will be determined variationally. The idea is to use the first term as the unperturbed Hamiltonian and the second one as the perturbation treated in the mean-field theory. Here we perform the change of variables   $u({\bf x}) = -\Delta h({\bf x})$ then $w({\bf x}) = \sqrt{\beta(\kappa({\bf x})+\lambda({\bf x}))}\,u({\bf x})$ to obtain an effective Hamiltonian in terms of the variable $w({\bf x})$ 
 \begin{eqnarray}
\fl\beta H'_{Hel} = {a^2\over 2}\sum_{\bf x} w^2({\bf x}) + \nonumber \\ \fl a^2 \sum_{{\bf x},{\bf x}',{\bf x}''}\overline\kappa({\bf x})\left[D_x^2 G_L({\bf x},{\bf x}') D_y^2 G_L({\bf x},{\bf x}'')  -
D_xD_y G_L({\bf x},{\bf x}') D_xD_y G_L({\bf x},{\bf x}'')\right]{w({\bf x}')w({\bf x}'')\over \sqrt{\kappa({\bf x}')+\lambda({\bf x}')}\sqrt{\kappa({\bf x}'')+\lambda({\bf x}'')}}\nonumber \\
 -{a^2\over 2}\sum_{\bf x} {\lambda({\bf x}) w^2({\bf x})\over \kappa({\bf x}) +\lambda({\bf x})}.
\end{eqnarray}
Keeping track of the term coming from the change of variables, we find the effective mean-field free energy as a functional of $\lambda$ is given by
\begin{equation}
\fl F_{MMF}(\lambda) = {TN\over 2}\left(
\ln\left(1+ {1\over 2} \left[\langle {\overline\kappa\over \kappa+\lambda}\rangle_0-\langle \overline\kappa\rangle_0\langle {1\over {\kappa+\lambda}}\rangle_0\right] -\langle {\lambda\over 
\kappa+\lambda}\rangle_0
\right) + \langle\ln\left[\beta (\kappa+\lambda)\right]\rangle_0\right),\label{pt2}
\end{equation}
and we note that $F_{MMF}(0)=F_{MMF}$ for the standard mean-field approximation given in Eq. (\ref{mmf1}). The mean-field theory as set up  provides an upper bound for the true free energy and thus we minimize $F_{MMF}(\lambda)$ with respect to $\lambda$. This gives a self consistent equation for $\lambda$ : $\ {\delta F_{MMF}(\lambda)/\delta\lambda({\bf x})} = 0$, which gives
\begin{equation}
\lambda({\bf x}) = {c\over 2}\left(\overline{\kappa}({\bf x}) -\langle \overline\kappa({\bf x})\rangle_0 \right) + (c-1)\kappa({\bf x})\label{lam1}
\end{equation}
with $c$ a constant given by
\begin{equation}
c= \left({1\over 2} \left[\langle {\overline\kappa\over \kappa+\lambda}\rangle_0-\langle \overline\kappa\rangle_0\langle {1\over {\kappa+\lambda}}\rangle_0\right] +\langle {\kappa\over 
\kappa+\lambda}\rangle_0
\right)^{-1};\label{c}
\end{equation}
The resulting self-consistent equation for $c$ obtained by substituting Eq. (\ref{lam1}) into Eq. (\ref{c})  actually yields the trivial relation equation $c=c$. This is simply due to the presence of a zero mode in the free energy, which turns out to be independent of $c$, and which is given by
\begin{equation}
F_{MMF}= {TN\over 2}\left[\ln(\beta) + \langle \ln\left(\kappa + {1\over 2}  \overline \kappa - {1\over 2} \langle \overline \kappa\rangle_0\right)\rangle_0\right].
\end{equation}
In addition, with this choice of $\lambda$ the perturbative correction is identically zero, corresponding to the vanishing of the first term in Eq. (\ref{pt2}), meaning that the variationally improved perturbation theory is also compatible with what is often known as {\em  self-consistent perturbation theory}, where the unperturbed Hamiltonian is self-consistently chosen so that the first order correction is zero \cite{dea01}. For the case of a two-component system, the membrane mean-field free energy per lipid patch is then given by 
\begin{equation}
\beta f_{mmf}(\phi) = {1\over 2}\phi \ln\left[\kappa_1 +{1\over 2}(1-\phi)(\overline\kappa_1-\overline \kappa_2)\right] + {1\over 2}(1-\phi)\ln\left[\kappa_2 -{1\over 2}\phi(\overline\kappa_1-\overline \kappa_2)\right]
\label{mfm2}.
\end{equation}
 We immediately see that an effective interaction exists between the two lipid species even when $\kappa_1 = \kappa_2$ as long as the Gaussian rigidities are not the same. This is in qualitative agreement with the exact two body result Eq. (\ref{2bexact}) found in \cite{lin11}. To understand the consequences of this result, consider the free energy difference per patch $\Delta f_{mmf}$ between a mixture and one that is phase separated into two components which is given by
 \begin{equation}
 \beta\Delta f_{mmf} = {1\over 2}\phi \ln\left[1 +{1\over 2\kappa_1}(1-\phi)(\overline\kappa_1-\overline \kappa_2)\right] + {1\over 2}(1-\phi)\ln\left[1 -{1\over 2\kappa_2}\phi(\overline\kappa_1-\overline \kappa_2)\right].
 \end{equation}
 Expanding this as a series in $\Delta\overline\kappa= \overline\kappa_1 -\overline\kappa_2$ we obtain
 \begin{equation}
 \beta\Delta f_{mmf} = {1\over 4}\Delta\overline\kappa ({1\over \kappa_1}-{1\over \kappa_2})\phi(1-\phi) -{1\over 16}\Delta\overline\kappa^2 \phi(1-\phi)\left[ {1-\phi \over \kappa_1^2}
 +{\phi \over \kappa_2^2}\right] +O(\Delta\overline\kappa^3).
 \end{equation}
 We see that the first term proportional to $\phi(1-\phi)$ is the effective two body Flory parameter, as given in Eq. (\ref{echif});its sign can be positive or negative. The second term is however always negative and favors mixing at small values of $\phi$ or $1-\phi$. Indeed it is this term that dominates when $\kappa_1=\kappa_2$.  
 
Once again the bounds of Eq. (\ref{bounds}) ensure that the variationally improved mean-field free energy is finite. As in the case of the ordinary mean-field approximation, in the absence of additional interactions between the lipids,  we find that the fluctuation-induced interactions are not sufficient to generate a phase separation as when $\chi_d=0$ we have verified numerically that the free energy remains convex.
 
The leading order correction due to the presence of variations in Gaussian rigidity is to map the problem onto one with no variations in the Gaussian bending rigidity but with an effective local bending rigidity given by
\begin{equation}
\kappa_{eff}({\bf x})=  \kappa({\bf x}) + {1\over 2}\left(\overline{\kappa}({\bf x}) -\langle \overline\kappa({\bf x})\rangle_0\right).
\end{equation}
This in turn leads to an effective bending rigidity for small Fourier modes given by
\begin{equation}
\kappa_e = \langle {1\over \kappa + {1\over 2}\left(\overline{\kappa} -\langle \overline\kappa\rangle_0\right)}\rangle_0^{-1}.\label{kev}
\end{equation}

\begin{figure}[t]
   \centering
  \includegraphics[width=14cm]{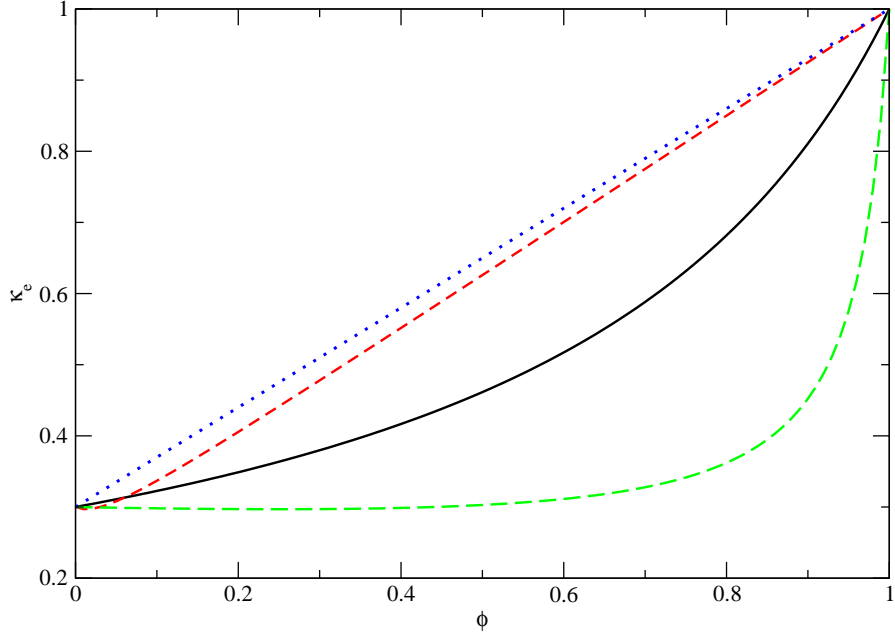}
  \caption{ Variationally improved mean-field theory prediction for the effective bending rigidity of a membrane in terms of its composition $\phi$ with $\kappa_1=1$ and $\kappa_2=0.3$ The solid black curve is the case where there are no interactions corresponding to $\chi_f=0$. The upper dashed curve (red -color on line) corresponds to the case where $\overline\kappa_1=-2\kappa_1=-2$ while $\overline\kappa_2=-0.1$. The upper dashed curve (green - color on line) corresponds to the other limiting case where $\overline\kappa_2=-2\kappa_2=-0.6$ while $\overline\kappa_1=-0.1$. The dotted (blue-collar online) line corresponds to the case where both Gaussian rigidities are minimal, i.e. $\overline\kappa=-2\kappa$.}.
\label{kappavar1}
\end{figure}

In the case where both Gaussian rigidities are minimal, $\overline \kappa=-2\kappa$, the effective rigidity has the very simple form $\kappa_e =\langle \kappa\rangle_0$. This is the straight-line behavior shown in Fig. (\ref{kappavar1}) (dotted line) for the case where $\kappa_1=1$ and $\kappa_2=0.3$. For the same bending ridigities, in  the same figure, we see the case where $\overline\kappa_1=-2$ while $\overline\kappa_2=-0.1$. This curve is lower than the noninteracting (solid black line) in a small region where the concentration of  the more rigid species $1$ is small. As the concentration of species 1 increases, the rigidity increases, crossing the noninteracting curve around $\phi=0.06$. The lower dashed line shows the case where $\overline\kappa_1=-0.1$ and $\overline\kappa_2=-0.6$. Here we see that the addition of a small amount of the phase $2$ to a pure membrane of the phase 1 dramatically reduces the effective rigidity of  the membrane. 
One should also note that the value $\langle\kappa\rangle_0$ can be shown to be an upper bound for the expression given in Eq. (\ref{kev}).

\section{Conclusions}

We have discussed how variations in the bending and Gaussian rigidity of the  components of simple model membranes induce effective fluctuation interactions due to their modification of height fluctuations. Apart from these, there are always interactions between the lipids composing membranes due to steric and van der Waals interactions. These interactions are present both when the membrane is flat, for instance if it is adhered to a flat surface, and when it is allowed to fluctuate. In this paper we have considered the additional interactions induced by the coupling of composition  to height fluctuations via composition dependent rigidities. In the case of tensionless membranes, we have exploited a simple transformation from the height variable to the reweighted mean curvature to demonstrate that variations in bending rigidity alone cannot induce height fluctuation mediated interactions between membrane components. When variations in the Gaussian rigidity $\overline\kappa$ are present, there are effective long range interactions, as first shown in \cite{gou93} at the pairwise level. A naive mean-field theory of  this system thus fails to describe the proper physics and indeed predicts erroneous results for the case where we have exact results. This is to be expected as the basic formulation of mean-field theory only treats the first term in the cumulant expansion, whereas fluctuation-induced interactions only show up in the  second order term of the cumulant expansion. However, by reformulating the theory in terms of the reweighted bending rigidity, the resulting theory contains all pairwise interactions in the first term of the cumulant  expansion. The corresponding mean-field approximation thus captures the basic fluctuation-induced interactions for this system, at least at the pairwise level. 

The resulting mean-field theory is characterized by an effective Flory parameter $\chi_f$ which depends on the bending rigidities of a two lipid system via Eq. (\ref{echif}). This result could in principle be used to estimate the difference in the Gaussian rigidities via an analysis of tensionless, or near tensionless, membrane fluctuations for membranes composed of lipid mixtures, by fluctuation mode analysis for example.  We emphasize that only relative differences can be measured, however this does present a step forward as previously only methods relying on topological changes in bilayer systems had been proposed to measure Gaussian  rigidities \cite{hu12}. 

The underlying model used is very idealized in the sense that the leaflets composing the bilayer are assumed to be symmetric, having the same composition on either side of the membrane - in this sense it is really a monolayer model. This condition can be relaxed by considering two coupled underlying lattices and taking the sum of the underlying bilayer bending energies. The effects of surface tension have also been ignored in this model. In practice, even if the system is not under any external constraints one should introduce a surface tension to fix the average area occupied by the lipids.  Finally, we note that most theoretical studies associate a bending rigidity with a lipid species as we have done here. However when one considers lattice based models, it is clear that rigidity is associated with lattice links rather than sites and should thus depend on the lipid at a site and its neighbors. The mean-field approach proposed here could be applied to such models, potentially giving rise to a richer behavior as well as more detailed comparison with experiments and numerical simulations, notably for the behavior of the effective bending rigidity \cite{imp}. 

The system studied here is rather special in that we have heavily exploited the exact solution for systems where $\overline \kappa$ is constant. For dielectric mixtures, one could presumably try to find  a similar strategy where one reformulates the field theory in such a away that pairwise  van der Waals interactions are treated exactly at the first order of the cumulant expansion. The development of a successful mean-field theory as a first method of studying the thermodynamics of systems dominated by fluctuation-induced interactions could be very useful to predict phases exhibited by such systems and consequently could guide both experimental and numerical studies.  

\noindent{\bf Acknowledgements} V.A.P. and R.P. acknowledge support by the U.S. Department of Energy, Office of Basic Energy Sciences, 
Division of Materials Sciences and Engineering under Award No. DE- SC0008176.

 \end{document}